\documentclass[conference,letterpaper]{IEEEtran}

\usepackage[utf8]{inputenc} 
\usepackage[T1]{fontenc}
\usepackage{url}
\usepackage{ifthen}
\usepackage{cite}
\usepackage[cmex10]{amsmath} 
\usepackage{amssymb,amsthm}
\usepackage{graphicx}

\ifCLASSOPTIONcompsoc
 \usepackage[caption=false,font=normalsize,labelfont=sf,textfont=sf]{subfig}
\else
 \usepackage[caption=false,font=footnotesize]{subfig}
\fi

\newtheorem{theorem}{Theorem}
\newtheorem{lemma}{Lemma}

\newtheorem{corollary}{Corollary}
\newtheorem{remark}{Remark}

\newtheorem{definition}{Definition}
\newtheorem{proposition}{Proposition}

\newcommand{\code}{\mathcal{C}}

\newcommand{\R}{\mathbb{R}} 
\newcommand{\Rd}{\mathbb{R}^d} 

\newcommand{\loss}[1]{\ell\left(#1\right)} 
\newcommand{\w}{\mathbf{w}} 
\newcommand{\xsubi}{\mathbf{x}_i} 
\newcommand{\ysubi}{y_i} 

\newcommand{\xj}[1]{\mathbf{x}_{#1}} 
\newcommand{\yj}[1]{y_{#1}} 

\newcommand{\watt}[1]{\mathbf{w}^{(#1)}} 
\newcommand{\gi}[1]{g_{#1}} 
\newcommand{\Di}[1]{D_{#1}} 
\newcommand{\Wi}[1]{W_{#1}} 
\newcommand{\Li}[1]{L_{#1}} 
\newcommand{\tildegij}[2]{\tilde{g}^{(#1)}_{#2}}

\newcommand{\Dsupi}[1]{D^{(#1)}} 
\newcommand{\gsupi}[1]{g^{(#1)}} 

\newcommand{\funct}[2]{f_{#1}\left(#2\right)} 
\newcommand{\cond}[1]{\textsf{cond}\left(#1\right)}

\newcommand{\bigOh}[1]{\mathcal{O}\left(#1\right)} 

\newcommand\blfootnote[1]{%
  \begingroup
  \renewcommand\thefootnote{}\footnote{#1}%
  \addtocounter{footnote}{-1}%
  \endgroup
}

\usepackage[inline]{trackchanges}
\addeditor{ozan}

\begin{document}
\title{Communication-Efficient Gradient Coding for Straggler Mitigation in Distributed Learning} 


\author{%
   \IEEEauthorblockN{Swanand Kadhe, O. Ozan Koyluoglu, and Kannan Ramchandran}
   \IEEEauthorblockA{Dept. of Electrical Engineering and Computer Sciences\\
                     University of California, Berkeley\\
                     Emails: \{swanand.kadhe,ozankoyluoglu,kannanr\}@berkeley.edu}
 }


\maketitle

\begin{abstract}
Distributed implementations of gradient-based methods, wherein a server distributes gradient computations across worker machines, need to overcome two limitations: delays caused by slow running machines called \emph{stragglers}, and communication overheads. Recently, Ye and Abbe [ICML 2018] proposed a coding-theoretic paradigm to characterize a fundamental trade-off between computation load per worker, communication overhead per worker, and straggler tolerance. However, proposed coding schemes suffer from heavy decoding complexity and poor numerical stability. In this paper, we develop a communication-efficient gradient coding framework to overcome these drawbacks. Our proposed framework enables using any linear code to design the encoding and decoding functions. When a particular code is used in this framework, its block-length determines the computation load, dimension determines the communication overhead, and minimum distance determines the straggler tolerance.
The flexibility of choosing a code allows us to gracefully trade-off the straggler threshold and communication overhead for smaller decoding complexity and higher numerical stability.
Further, we show that using a maximum distance separable (MDS) code generated by a random Gaussian matrix in our framework yields a gradient code that is optimal with respect to the trade-off and, in addition, satisfies stronger guarantees on numerical stability as compared to the previously proposed schemes. Finally, we evaluate our proposed framework on Amazon EC2 and demonstrate that it reduces the average iteration time by 16\% as compared to prior gradient coding schemes.
\end{abstract}


\blfootnote{This work is supported in part by National Science Foundation grants CCF- 1748585 and CNS-1748692.
}

\section{Introduction}
\label{sec:intro}
The scale of training datasets and model parameters in real-world machine learning applications is continuously growing. Therefore, it has become crucial to implement learning algorithms in a distributed fashion. A commonly used distributed learning framework is data parallelism, in which large-scale datasets are distributed over multiple {\it worker machines} for parallel processing in order to speed up computation.

Gradient descent based algorithms form an important class of learning algorithms, and are popular in practice. 
In a typical distributed set up using a gradient descent based algorithm,  
each worker machine computes {\it partial gradients} using their local data batches, and sends them to a parameter server which aggregates them to update the model parameters. 
Any distributed implementation of gradient descent (in a synchronous setting) needs to address the following two  challenges: 
delays caused by stragglers and communication overheads.

The latency performance of every iteration of a synchronous distributed gradient descent algorithm is determined by {\it stragglers} -- workers that are slowed down due to unpredictable factors such as network latency, hardware failures, etc.~\cite{Hoefler:10:noise,Dean:13:tail}.
Using coding-theoretic ideas to mitigate stragglers has  gained significant research attention, see, e.g.,~\cite{Lee:18,Dutta:16,Avestimehr:18:coded-matrix,Aktas:17} for distributed computing, and~\cite{Tandon:17,Halbawi:17,Raviv:18,Avestimehr:17,CharlesP:17,CharlesP:18:ISIT,Maity:19:LDPC,Diggavi:19,Avestimehr:18:regression,Bitar:19:stochastic-GC,Shroff:19,KadheKK:19} for distributed learning.  

 The gains due to parallelization are also bottlenecked in practice by heavy communication overheads between workers and the parameter server. 
 This is especially the case for modern deep learning applications using models with millions of parameters (e.g., ResNet~\cite{ResNet:16}).
 One approach to reduce communication overheads is to compress and quantize gradients, see, e.g., ~\cite{Wen:17:TernGrad,Alistarh:17:QSGD,Alistarh:18:sparsification,Avestimehr:18:gradiveq}. 
Coding theoretic ideas have also been proposed to trade-off computation for saving communication, see, e.g.,~\cite{Avestimehr:17:heterogeneous-communication-coding,Avestimehr:18:communication-computing,Avestimehr:18:compressed-computing,Avestimehr:19:coded-reduce}.

Recently, Ye and Abbe~\cite{YeAbbe:18i,YeAbbe:18} proposed a gradient coding paradigm to mitigate stragglers in distributed gradient aggregation and simultaneously reduce the communication overhead for workers. We refer to this paradigm as communication-efficient gradient coding.
The setup consists of $n$ worker machines and a parameter server. Training {samples} are partitioned into $k$ parts, and every worker is assigned $l$ of the $k$ parts. Every worker first computes the partial gradients each of length $d$ on its assigned {samples}, then encodes them into a vector of length $d/m$, and returns the result to the server. The data placement and encoding should ensure that the server can recover the sum of gradients even if any $s$ workers straggle. Note that $l$ specifies the computation load per worker, $m$ specifies the communication saving, and $s$ specifies the straggler tolerance. 

The communication-efficient gradient coding paradigm characterizes a three dimensional trade-off between $l$, $m$, and $s$. 
In particular, Ye and Abbe~\cite{YeAbbe:18i,YeAbbe:18} showed that any coding scheme with communication saving $m$ and straggler tolerance $s$ must incur the computation load of $l \geq k(s + m)/n$. This implies that the higher the straggler tolerance and/or communication saving, the larger is the computation load per worker.
They also presented two schemes that achieve the smallest computation load $l$ for a given communication saving $m$ and straggler tolerance $s$---one based on recursive polynomials and another based on random Gaussian matrices. 

The main drawback of the schemes in~\cite{YeAbbe:18i,YeAbbe:18} is that, to recover the gradient sum, the server needs to invert a matrix of size $n-s$.
This results in a decoding complexity of $\bigOh{n^3}$.  
In addition, the proposed schemes demonstrate poor numerical stability. Specifically, the authors observe that the numerical stability of the recursive polynomials based scheme quickly deteriorates as $n$ becomes larger than $20$, and the scheme becomes unstable for $n > 25$. The Gaussian-random-matrix based scheme is observed to be numerically stable for $n\leq 30$, however, its numerical stability performance for larger number of machines remains unclear. {The goal of this paper is to design schemes that overcome these drawbacks.}

{\bf Our Contributions:} We present a communication-efficient gradient coding framework which enables gracefully trading off the straggler threshold and communication saving for higher numerical stability and smaller decoding complexity. Our proposed framework uses the {\it fractional repetition} scheme to redundantly assign training data across workers, similar to~\cite{Tandon:17}, wherein the workers are partitioned into several groups and all the workers in a given group are assigned {to a subset of samples}. We demonstrate that the fractional repetition placement enables one to use any linear code to design the encoding and decoding functions
We refer to this framework as {\bf Comm}unication-Efficient {\bf F}ractional {\bf R}epetition-based {\bf G}radient {\bf C}oding {\it (CommFR-GC)}.

This flexibility of CommFR-GC to use any linear code enables us to choose a suitable code to reduce decoding complexity and achieve strong numerical stability guarantees. As a case study, we demonstrate that using a low-density parity check (LDPC) code in CommFR-FC yields a gradient code with decoding complexity linear in the number of workers. 
Next, we show that using a maximum distance separable (MDS) code generated by a random Gaussian matrix in CommFR-GC yields an optimal gradient code with stronger guarantees on numerical stability as compared to the schemes in~\cite{YeAbbe:18}.  
Specifically, we consider the straggler threshold under a given numerical stability requirement as in~\cite{YeAbbe:18}, and prove that the CommFR-GC with an MDS code achieves higher straggler thresholds for a wide range of parameters.
Finally, we evaluate our proposed framework on Amazon EC2 and demonstrate that it reduces the average iteration time by 16\% as compared to prior gradient coding schemes.

{
{\bf Related Work:} In~\cite{Avestimehr:19:coded-reduce}, the authors presented a joint design of data allocation, communication strategy, and gradient coding that mitigates stragglers as well as reduces bandwidth congestion at the parameter server. Our focus on the other hand is on reducing communication overhead at workers along with straggler mitigation. In~\cite{Maity:19:LDPC}, the authors proposed a straggler-robust gradient descent scheme for linear models that uses LDPC codes to encode the second-moment of the data. In contrast, our proposed coding framework, similar to~\cite{YeAbbe:18}, can be used for any distributed learning problem. In~\cite{Cadambe:19,Heidarzadeh:19,Grover:19}, the authors develop numerically stable coding schemes for  polynomial computations and distributed matrix multiplication. Here, we focus on designing numerically stable gradient coding schemes.
}

\section{Problem Setup}
\label{sec:problem-setup}

\subsection{Distributed Training}
\label{sec:training}
The process of learning the parameters $\w\in\Rd$ of a model given a dataset $D = \{(\xsubi,\ysubi)\}_{i=1}^{M}$ of $M$ samples, where $\xsubi \in \Rd$ and $\ysubi\in \R$, can be cast as the {\it empirical risk minimization} (ERM) problem:  $\min_{\w} \frac{1}{M}\sum_{i=1}^{M} \loss{\xsubi,\ysubi;\w},$
where $\loss{\xsubi,\ysubi;\w}$ is a loss function that measures the accuracy of the prediction made by $\w$ on  $(\xsubi,\ysubi)$. 

In distributed settings, a popular method to approximately solve the ERM 
is mini-batch stochastic gradient descent (SGD). In every iteration of mini-batch SGD, a (possibly random) subset $S_t$ of $B$ samples is chosen and the model is updated as \mbox{$\watt{t+1} = \watt{t} - \frac{\alpha_t}{B}\sum_{i\in S_t} \nabla\loss{\xsubi,\ysubi;\watt{t}}$,} where $\alpha_t$ is the learning rate at iteration $t$.
In the remainder of the paper, we focus our attention to a given iteration $t$, and fix a batch of $B$ samples. 
We omit the explicit dependence on the iteration $t$ hereafter, since our focus is on a given iteration. 

\subsection{Gradient Coding}
\label{sec:grad-coding}
In this section, we describe the communication-efficient gradient coding setup from~\cite{YeAbbe:18i,YeAbbe:18}. Consider a distributed master-worker setting consisting of $n$ worker machines $\Wi{1},\Wi{2},\ldots,\Wi{n}$, and a parameter server. 
The mini-batch of samples is partitioned into $k$ subsets of equal size,
denoted as $\Di{1},\Di{2},\ldots,\Di{k}$. Define the gradient vector of the partial data $\Di{i}$, called {\it partial gradient}, as \mbox{$\gi{i} := \sum_{\mathbf{x}_j,\mathbf{y}_j\in\Di{i}} \nabla \loss{\xj{j},\yj{j};\w}$.} 
A communication-efficient gradient code is parameterized by three (non-negative) integers: its per worker computation load $l$, per worker communication saving $m$, and straggler tolerance $s$. 
In particular, given $n$ and $k$, a gradient code is said to achieve a triple $(l,m,s)$, satisfying \mbox{$1\leq l\leq k$,} \mbox{$0\leq s \leq n-1$,} and \mbox{$m\geq 1$,} if there exist
\begin{enumerate}
\item a {\it placement scheme} that assigns $l$ datasets to each worker;
\item an {\it encoding scheme} that allows every worker $\Wi{i}$, \mbox{$1\leq i\leq n$,} to encode its $l$ partial gradients $\{\gi{i_1}, \gi{i_2}, \ldots, \gi{i_l}\}$, total of a $dl$ dimensional vector, to a $\lceil d/m \rceil$-dimensional vector; and
\item a {\it decoding scheme} that allows the parameter server to decode the sum of gradients $\sum_{i=1}^k\gi{i}$ from any subset of $n-s$ workers.
\end{enumerate}

\noindent 
As in~\cite{YeAbbe:18i}, we restrict our attention to linear coding schemes
for the sake of low complexity.

It is shown in~\cite{YeAbbe:18i} that, given $n$ and $k$, a triple $(l, m, s)$ is achievable if and only if 
\begin{equation}
\label{eq:lower-bound}
\frac{l}{k} \geq \frac{s+m}{n}.
\end{equation}
This essentially shows that the fractional computation load $l/k$ at each worker must increase with the straggler tolerance $s$ and the communication savings $m$. Further, one can observe that for $m=1$, the result reduces to the case of gradient coding without any communication saving considered in~\cite{Tandon:17,Halbawi:17,Raviv:18}.

{\bf Coding Schemes in~\cite{YeAbbe:18}}:
The schemes proposed in~\cite{YeAbbe:18} use cyclic placement, which assumes $k = n$, and assigns to the $i$-th worker the following $l$ datasets $\{D_i, D_{(i + 1) \mod n}, \ldots, D_{(i + l - 1) \mod n}\}$. 
Encoding uses two matrices $B\in\R^{mn \times (n-s)}$ and $V\in\R^{(n-s)\times n}$. The matrices are chosen such that any $(n-s)\times(n-s)$ sub-matrix of $V$ is non-singular, and a specific set of the rows of $B$ is orthogonal to $V$. (See Appendix~\ref{app:YeAbbe} for details.)
To recover the gradient sum, the server effectively needs to invert an $(n-s)\times(n-s)$ matrix $V_T$, where $T\subseteq[n]$ is the set of non-straggling workers and and $V_T$ is the $(n-s)\times |T|$ sub-matrix of $V$ corresponding to the columns indexed by $T$.
The authors present two coding schemes---when $V$ is a Vandermonde matrix, and when $V$ is a random Gaussian matrix. 

The main drawback of the schemes in~\cite{YeAbbe:18} is that, in the decoding phase, the server needs to invert an $(n-s) \times (n-s)$ matrix. This not only increases the decoding complexity, but, more importantly, also results in poor numerical stability. 
In the next section, we present a novel framework to construct gradient coding schemes 
which can gracefully trade off their straggler threshold and communication saving for higher numerical stability and/or smaller decoding complexity. 

\section{Communication-Efficient Fractional Repetition Based Gradient Coding (CommFR-GC)}
\label{sec:proposed-framework}
We present a coding framework that utilizes any arbitrary linear code (over $\mathbb{R}$) to design encoding and decoding schemes. 
We begin with a brief review of linear codes over $\mathbb{R}$ (see, e.g.,~\cite{Marshall:84}). A linear code $\code$ of block-length $N$ and dimension $K$ is a $K$-dimensional subspace of $\R^N$. The minimum distance $\delta$ of $\code$ is $\min\{d_H(x, y) : x, y \in \code, x\ne y\}$, where $d_H$ denotes the Hamming distance \mbox{$d_H(x, y) =|\{1\leq i\leq N \mid x_i\ne y_i\}|$.} 
Such a code is denoted as an $[N,K,\delta]$ code. The well-known upper bound on the minimum distance, called Singleton bound, is given as $\delta \leq N - K + 1$, and codes which attain this bound with equality are referred to as Maximum Distance Separable (MDS) codes.

An $[N, K]$ linear code $\code$ can be described using a generator matrix $G\in\R^{K\times N}$ of rank $K$, whose rows form a basis of $\code$.  Alternatively, it is possible to describe a linear code $\code$ using a parity-check matrix $H \in\R^{(N-K)\times N}$ of rank $N-K$ such that $\code$ is the null space of $H$.
A generator matrix leads to a {\it systematic} encoding, if for each $\mathbf{x}\in\R^K$, the  codeword $\mathbf{c} = \mathbf{x}G$ contains $\mathbf{x}$ on some $k$ coordinates.

\subsection{CommFR-GC Framework}
\label{sec:proposed-scheme}
Consider the parameters $n$, $k$, $N$, and $K$ such that $N\mid n$ and $n \mid kN$. Let $l = kN/n$. Further, let $p = n/N = k/l$. We propose a  framework based on fractional repetition placement scheme, considered in the first gradient coding work~\cite{Tandon:17}. This placement enables us to use of any $[N, K, \delta]$ code to construct a communication-efficient gradient code. We refer to the framework as Communication-efficient Fractional Repetition Gradient Coding (CommFR-GC).
The CommFR-GC coding framework consists of three components: placement, encoding, and decoding described in the following.  

\noindent{\bf 1. Fractional Repetition Placement Scheme}:
The $k$ datasets $\{\Di{1},\cdots,\Di{k}\}$ are assigned to $n$ workers as follows. 
\begin{enumerate}
    \item Partition $n$ workers into $p$ groups each of size $N$, denoted as $\Li{1},\ldots,\Li{p}$.
    \item Each of the $N$ workers in group $\Li{i}$ stores the following $l$ datasets 
        $ 
        \Dsupi{i} = \left\{\Di{(i-1)l+1},\Di{(i-1)l+2},\cdots,\Di{il}\right\}
        $.
\end{enumerate}

\noindent{\bf 2. Encoding Scheme (Worker Side)}:
Let $G\in\mathbb{R}^{K\times N}$ be a generator matrix of an $[N, K, \delta]$ code $\code$. 
\begin{enumerate}
    \item Each worker first computes the sum of partial gradients for all datasets assigned to it. In particular, for $1\leq i\leq p$, every worker in group $\Li{i}$ computes 
   \begin{equation}
   \label{eq:group-gradient}
       \gsupi{i} = \sum_{u\in\Dsupi{i}}\gi{u}.
   \end{equation}
   
    \item Each worker in group $i$, $1\leq i\leq p$, appends $\gsupi{i}$ with $(K\lceil d/K \rceil - d)$ zeros.
    
    \item For $1\leq i\leq p$, $1\leq j\leq N$, the $j$-th worker in group $\Li{i}$ computes 
    \begin{equation}
    \label{eq:coded-gradient}
        \tilde{g}^{(i)}_j = \gsupi{i}_{mat}G_j,
    \end{equation} 
    where $G_j$ is the $j$-th column of $G$, and $\gsupi{i}_{mat}$ is a $\lceil d/K \rceil \times K$ matrix obtained by arranging $\gsupi{i}$ as
    \begin{equation}
    \label{eq:gradient-matrix}
        \gsupi{i}_{mat} = 
        \begin{bmatrix}
            \gsupi{i}(1) & 
            \cdots & 
            \gsupi{i}((K-1)\lceil d/K \rceil+1)\\
            \gsupi{i}(2) & 
            \cdots & 
            \gsupi{i}((K-1)\lceil d/K \rceil+2)\\
            \vdots    & 
            \ddots & 
            \vdots\\
            \gsupi{i}(\lceil d/K \rceil) & 
            \cdots & 
            \gsupi{i}(K\lceil d/K \rceil)
        \end{bmatrix}.
    \end{equation}
    
    \item For $1\leq i\leq p$, the $j$-th worker in group $\Li{i}$ sends $\tilde{g}^{(i)}_j$ (of length $\lceil d/K \rceil$) to the parameter server.
\end{enumerate}

\noindent{\bf 3. Decoding Scheme (Server Side):} 
The parameter server waits for the first $t$ $(\geq m)$ workers  
from each group, and decodes the gradient sum as follows. (We characterize the exact value of $t$ in Theorem~\ref{thm:generic-threshold}.) 
\begin{enumerate}
    \item For a group $i$, 
    let $i_1$, $i_2$, $\cdots$, $i_t$ be the first $t$ workers to finish. From these $t$ workers, the master receives 
    $\begin{bmatrix}
    \tildegij{i}{i_1} & \tildegij{i}{i_2} & \cdots & \tildegij{i}{i_t}
    \end{bmatrix}$. 
    \item For each group $1\leq i\leq p$, the master {decodes} $\gsupi{i}_{mat}$ by solving 
    \begin{equation}
        \label{eq:decoding-a-group}
        \begin{bmatrix}
        \tildegij{i}{i_1} & \tildegij{i}{i_2} & \cdots & \tildegij{i}{i_t}
        \end{bmatrix}
        = \gsupi{i}_{mat}
        G^{(i)},
    \end{equation}
    where $G^{(i)}$ is the $K\times t$ sub-matrix of $G$ consisting of the columns $i_1$, $i_2$, $\cdots$, $i_t$. 
    \item For each group $1\leq i\leq p$, the master obtains $\gsupi{i}$ by rearranging the entries of $\gsupi{i}_{mat}$ as a vector  and removing the last $(K\lceil d/K \rceil - d)$ zeros (see~\eqref{eq:gradient-matrix}).
    \item From $\gsupi{1}$, $\gsupi{2}$, $\cdots$, $\gsupi{p}$, the master computes the gradient sum as $g = \sum_{i=1}^{p}\gsupi{i}$. It is straightforward to verify that $g = \sum_{j = 1}^k g_j$ from~\eqref{eq:group-gradient}.
\end{enumerate}

\begin{figure}[!t]
\begin{center}
\includegraphics[scale = 0.25]{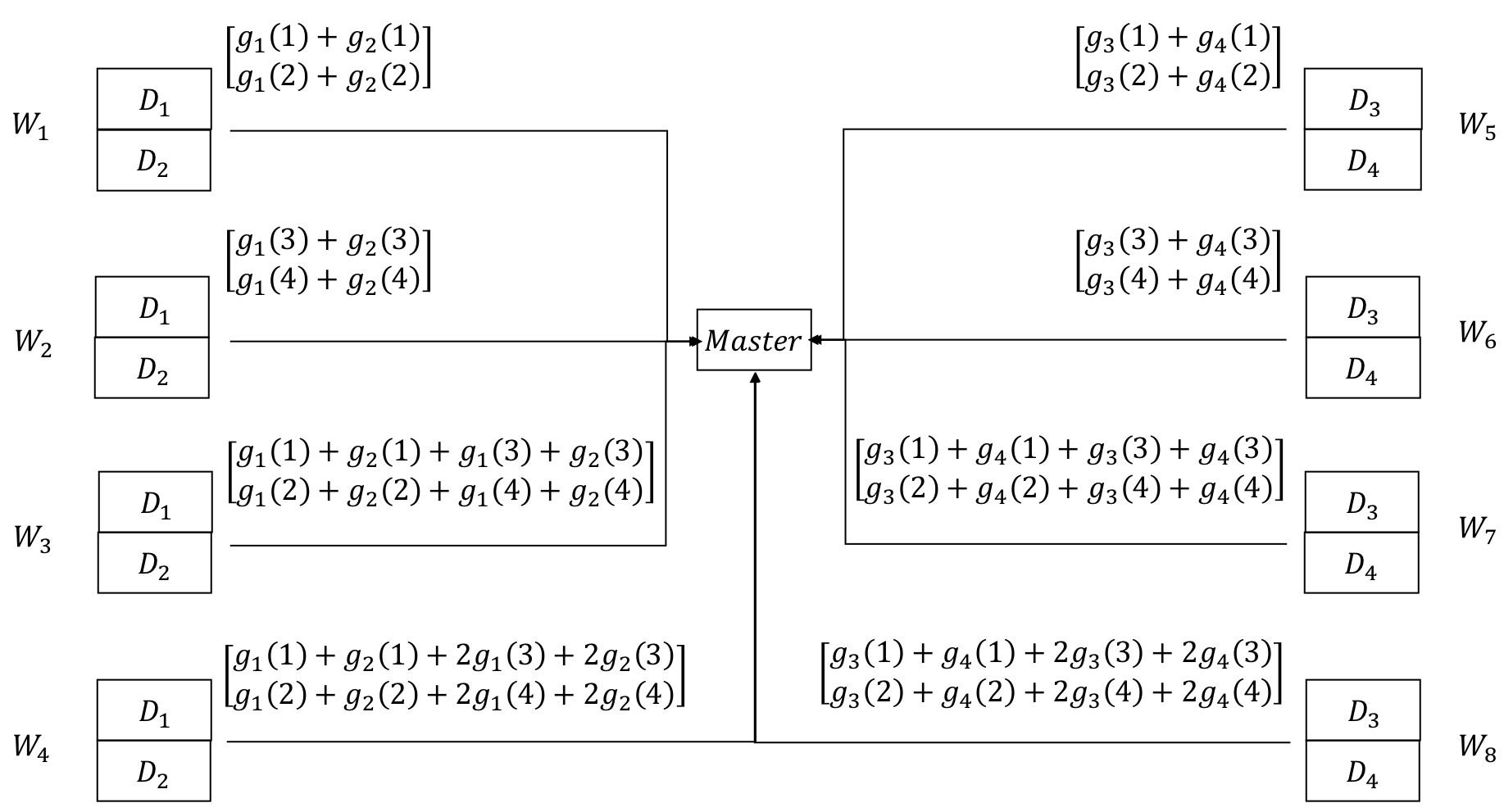}
\caption{
Example of CommFR-GC for $n = 8$ workers, $k = 4$ datasets, and gradient-length $d = 4$. We use an $[N =4, K = 2, \delta = 3]$ MDS code with the generator matrix in~\eqref{eq:example-G}.
This yields a gradient code with   computation load $l = 2$, communication saving $m = 2$, and straggler threshold $s = 2$.
}
\label{fig:code-ex}
\end{center}
\end{figure}

\noindent {\bf Example:} Consider $n = 8$ workers, $k = 4$ datasets, and gradient-length $d = 4$. We use an $[N=4, K=2, \delta=3]$ MDS code (over $\R$) with the following generator matrix:
\begin{equation}
\label{eq:example-G}
G = \begin{bmatrix}
1 & 0 & 1 & 1\\
0 & 1 & 1 & 2
\end{bmatrix}.
\end{equation} 
Note that $l = kN/n = 2$. 

\noindent{\it Fractional repetition placement scheme:} First four workers form group 1 and store $\Di{1},\Di{2}$, whereas the last four workers form group 2 and store $\Di{3},\Di{4}$ (see Fig.~\ref{fig:code-ex}). 

\noindent{\it Encoding scheme:} Each worker in group 1 (resp. group 2) computes $\gsupi{1} = \gi{1}+\gi{2}$ (resp. $\gsupi{2} = \gi{3}+\gi{4}$). 
For $1\leq j\leq 4$, the $j$-th worker in group $i$ computes, 
$$
\tildegij{i}{j} = \begin{bmatrix}
\gsupi{i}(1) & \gsupi{i}(3)\\
\gsupi{i}(2) & \gsupi{i}(4)
\end{bmatrix}G_j,
$$
where $G_j$ is the $j$-th column of $G$ (see Fig.~\ref{fig:code-ex}).

\noindent {\it Decoding:} Since any two columns of $G$ are independent, we can see that, even if any two workers straggle in each group $\Li{1}$ and $\Li{2}$, the master can recover $\gsupi{1} = \gi{1}+\gi{2}$ from non-stragglers in $\Li{1}$ and $\gsupi{2} = \gi{3}+\gi{4}$ from non-stragglers in $\Li{2}$. Thus, in the worst case, the scheme can tolerate any two stragglers.

\subsection{Trade-Off Characterization}
\label{sec:trade-off-proposed-scheme}
Here, we characterize the trade-off between the computation load, communication saving, and straggler threshold for the CommFR-GC framework as a function of the underlying code.
\begin{theorem}
\label{thm:generic-threshold}
Given $n$, $k$, $N$, and $K$ such that $N\mid n$ and $n\mid kN$, a CommFR-GC using an $[N, K, \delta]$ code achieves the triple $\left(l = kN/n, m =K, s = \delta-1\right)$. 
\end{theorem}
\begin{IEEEproof}
By construction, each worker computes $l = kN/n$ partial gradients, and communicates a vector of length $\lceil d/K \lceil$. 
To see that the straggler threshold is $\delta - 1$, note that, for a linear code with block-length $N$ and minimum distance $\delta$, it is always possible to solve~\eqref{eq:decoding-a-group} for any $t\geq N-\delta+1$. Thus, every group can tolerate up to $\delta-1$ stragglers, which results in the straggler threshold of $s = \delta - 1$.  
\end{IEEEproof}

\begin{corollary}
\label{cor:MDS-achieves-threshold}
Given $n$, $k$, $N$, and $K$ such that $N\mid n$ and $n\mid kN$, a CommFR-GC scheme using an $[N, K, N-K+1]$ MDS code is optimal with respect to~\eqref{eq:lower-bound}.
\end{corollary}

\begin{remark}
\label{rem:relation-to-FR}
CommFR-GC with an $[N, 1, N]$ repetition {code} reduces to the fractional repetition gradient code in~\cite{Tandon:17}. 
\end{remark}

\section{Analysis}
\label{sec:analysis}
We analyze how the choice of the code in CommFR-GC impacts the decoding complexity and numerical stability. 

\subsection{Decoding Complexity}
\label{sec:decoding-complexity}
First, we characterize the decoding complexity of an optimal CommFR-GC scheme that uses a systematic MDS code.

\begin{proposition}
\label{thm:decoding-complexity-MDS}
Using a systematic $[m+s, m, s+1]$ MDS code in the CommFR-GC framework yields an optimal gradient code with decoding complexity  $\bigOh{\frac{(\min\{m,s\})^3n}{\max\{m,s\}}}$.
\end{proposition}
\begin{IEEEproof}
The optimality with respect to~\eqref{eq:lower-bound} is immediate. The decoding complexity follows from the fact that to decode the code one needs to invert a matrix of size $t\times t$, where $t = \min\{m,s\}$, for each group, and the number of groups is $n/(m+s)$.
\end{IEEEproof}

\begin{remark}
\label{rem:decoding-complexity}
We compare the complexity with the schemes in~\cite{YeAbbe:18}. 
We consider the regime $s = \bigOh{n}$, as the number of stragglers in practical systems is typically a small percentage of the number of workers (see, e.g.,~\cite{Tandon:17,Gupta:oversketch:18,Gupta:OSN:19}). 
When
$m = \bigOh{n}$, CommFR-GC with an MDS code has $\bigOh{n^3}$ decoding complexity, which is same as that of the schemes in~\cite{YeAbbe:18}. 
On the other hand, CommFR-GC with an MDS code achieves smaller decoding complexity when $m$ is sub-linear in $n$. In other words, when $m = \bigOh{n^{\alpha}}$ for some $\alpha < 1$, then the decoding complexity is $\bigOh{n^{3\alpha}}$. 
\end{remark}

Next, we demonstrate that is possible to achieve linear decoding complexity (i.e., $\bigOh{n}$) by choosing a suitable code. 
In particular, we consider the well-known class of computationally efficient codes, called low-density parity check (LDPC) codes (see, e.g.,~\cite{Urbanke:08}). Note that our focus here is on the regime $s = m = \bigOh{n}$ with sufficiently large $n$. This regime is of interest for {\it serverless systems}~\cite{Jonas:17}, which can invoke several thousands of workers and suffer from significantly large number of stragglers~\cite{Gupta:oversketch:18,Gupta:OSN:19}. 

An $[N, K]$ LDPC code is defined using a sparse binary parity-check matrix $H$ of size $(N-K) \times N$, 
which is typically chosen randomly from an ensemble. 
LDPC codes admit an iterative decoder, called peeling decoder, that can recover a random subset of erasures with complexity $\bigOh{N}$.


As a case study, we consider a specific LDPC ensemble $\code(d_c,d_v)$ defined by the random binary parity-check matrix 
with each row having $d_c$ ones and each column having $d_v$ ones.
Let $p*$ be the largest $\epsilon \in (0,1)$ such that $\epsilon(1 - (1 - x)^{d_c -1})^{d_v - 1} < x $ for all $x \in (0,1]$. Note that $p*$ is referred to as the threshold associated with $(d_c, d_v)$. Let each coordinate of a codeword $\mathbf{c}\in\code$ be independently erased with some probability $p$. It can be shown that for sufficiently large $n$, for any $p \leq p*$, the peeling decoder recovers the codeword with high probability~\cite[Chapter 3]{Urbanke:08}. Utilizing this result, we state the following proposition.

\begin{proposition}
\label{thm:decoding-complexity-LDPC}
Consider the case $m = s = O(n)$, and a CommFR-GC using an $[m+s, m]$ LDPC code chosen from the $\code(d_c,d_v)$ ensemble. When each worker straggles independently with probability $p \leq p*$, the server can recover the gradient sum with high probability for sufficiently large $n$, with decoding complexity  $\bigOh{n}$.
\end{proposition}

\begin{remark}
\label{rem:decoding-complexity-LDPC}
As an example, let $n = 30000$ and $m = s = 5000$. Again, we note that this regime is applicable in serverless systems that usually invoke tens of thousands of workers~\cite{Jonas:17,Gupta:oversketch:18,Gupta:OSN:19}. We consider a CommFR-GC with a $[10000,5000]$ LDPC code chosen from the ensemble $\code(3,6)$. For this case, it is possible to show that $p* = 0.4294$~\cite[Chapter 3]{Urbanke:08}. Using a peeling decoder will allow the server to recover, with high probability, the gradient sum from a random set of $4249$ stragglers in each group with decoding complexity $\bigOh{10000}$. On the other hand, using a $[10000,5000]$ MDS code in CommFR-GC allows the server to recover from any set of $5000$ stragglers in each group with decoding complexity $\bigOh{10000^3}$. This demonstrates how CommFR-GC allows one to back off from the optimal straggler threshold to achieve smaller decoding complexity. In fact, using an LDPC code from a capacity-achieving ensemble will yield a scheme with an optimal straggler threshold, but only for applications in which stragglers are random.
\end{remark}

\subsection{Numerical Stability}
\label{sec:numerical-stability}
In this section, we demonstrate that CommFR-GC using a Gaussian random matrix as a generator matrix achieves significantly better numerical stability compared to the Gaussian matrix based scheme in~\cite{YeAbbe:18}.
To quantify the numerical stability, we consider the straggler threshold achievable under numerical stability parameter $\kappa$ as follows.

\begin{definition}\cite{YeAbbe:18}
\label{def:kappa-achievable-triple}
Given $n$ and $k$, a coding scheme designed to achieve a triple $(l,m,s)$ is said to have straggler tolerance $s_{\kappa}$ under numerical stability constraint $\kappa$, if the scheme can tolerate any $s_{\kappa}$ stragglers such that the condition number of any matrix involved in the decoding is upper bounded by $\kappa$.
\end{definition}


To aid our analysis, we define a function $\funct{s,m,\kappa}{t}$ for given $m$, $s$, $m\leq t\leq m+s$, and $\kappa$ as follows:
\begin{equation}
\label{eq:function-f}
\funct{s,m,\kappa}{t} = \frac{1}{\sqrt{2\pi}}\binom{m+s}{t}\left(\frac{Ct}{\kappa(t-m+1)}\right)^{t-m+1},
\end{equation}
where $C\leq 6.414$ is a universal positive constant.
Now, it is straightforward to verify that, for any
\begin{IEEEeqnarray}{l}
\label{eq:kappa-lower-bound}
    \kappa > \max\left\{\left(\frac{1}{\epsilon\sqrt{2\pi}}\right)^{1/(s+1)}\left(\frac{C(m+s)}{s+1}\right),\frac{Cs}{2}\right\}
\end{IEEEeqnarray}
the function $f_{s,m,\kappa}(t)$ is monotonically decreasing in $t$, and $f_{s,m,\kappa}(m+s)\leq \epsilon$. Therefore, for any $\kappa$ satisfying~\eqref{eq:kappa-lower-bound}, there exists an integer $m\leq t\leq m+s$ such that $f_{s,m,\kappa}(t)\leq \epsilon$. This allows us to analyze the straggler threshold of the proposed scheme under the numerical stability constraint $\kappa$ in terms of $f_{s,m,\kappa}(\cdot)$ as follows.

\begin{theorem}
\label{thm:straggler-threshold-kappa}
Consider the CommFR-GC scheme using an $[m+s,m]$ code defined by a generator matrix whose the elements are chosen as i.i.d. standard normal. For any $0 < \epsilon < 1$, and any $\kappa$ satisfying~\eqref{eq:kappa-lower-bound},
it holds, with probability at least $ 1 - \epsilon$, that $s_{\kappa} \geq s + m - t^*$, where $t^*$ is the smallest integer between $m$ and $s+m$ such that $\funct{m+s,m,\kappa}{t^*}\leq\epsilon$. 
\end{theorem}
\begin{IEEEproof}
See Appendix~\ref{app:proof-numerical-stability}.
 \end{IEEEproof}

\begin{table}[!t]
    \renewcommand{\arraystretch}{1.1}
        \caption{Straggler thresholds $s_{\kappa}^{YA}$ for the scheme in~\cite{YeAbbe:18} and $s_{\kappa}$ for CommFR-GC for various $n$, $s$, and $m$, when $k = n$, $\kappa = 1000$ and $\epsilon = 10^{-3}$. 
        }
        \vspace{-2mm}
    \centering
    \begin{tabular}{|c|c|c|c|c|}
         \hline
          $n$ & $s$ & $m$ & $s_{\kappa}^{YA}$ & $s_{\kappa}$\\
          \hline
          \hline
          60 & 3 & 2 & 0 & 2\\
          \hline 
          60 & 8 & 2 & 2 & 6\\
          \hline
          60 & 13 & 2 & 6 & 11\\
          \hline
          60 & 3 & 12 & 0 & 1\\
          \hline
          60 & 8 & 12 & 2 & 4\\
          \hline
          60 & 13 & 12 & 2 & 4\\
          \hline
          \hline
          1000 & 40 & 10 & 8 & 32\\
          \hline
          1000 & 90 & 10 & 29 & 78\\
          \hline
          1000 & 190 & 10 & 85 & 172\\
          \hline
          1000 & 40 & 210 & 8 & 16\\
          \hline
          1000 & 90 & 210 & 29 & 48\\
          \hline
          1000 & 190 & 210 & 85 & 121\\
          \hline
    \end{tabular}
    \label{tab:threshold-comparison}
    \vspace{-4mm}
\end{table}

\begin{figure}[!t]
    \centering
         \centering
        \includegraphics[scale=0.3]{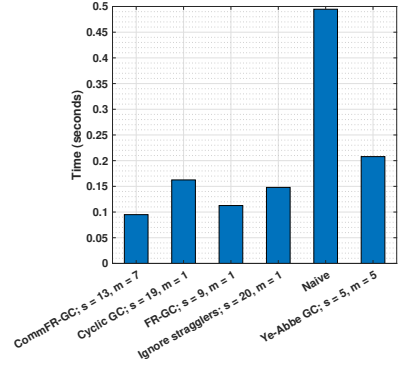} 
        \caption{Average time per iteration for $n = 60$ \texttt{t2.micro} worker instances on EC2, with gradient-length $d=241915$ for various coding schemes.}
        \label{fig:avg-time-n60}
    \vspace{0mm}
\end{figure}

 \begin{figure}[!t]
    \centering
        \centering
        \includegraphics[scale=0.3]{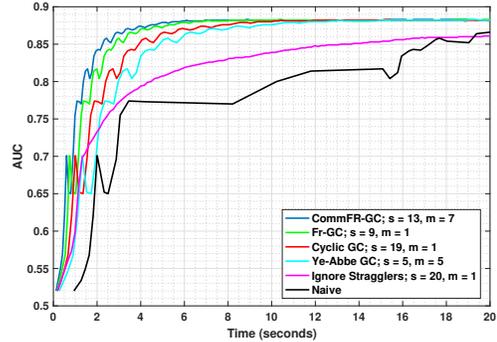} 
        \caption{AUC vs no. of iterations for $n = 60$ \texttt{t2.micro} worker instances on EC2, with gradient-length $d=241915$ for various coding schemes.}
         \label{fig:AUC-n60}
         \vspace{-4mm}
\end{figure}

{\bf Comparison with the Ye-Abbe scheme~\cite{YeAbbe:18}:} 
It is not hard to verify that the straggler threshold with numerical stability constraint for the Ye-Abbe  scheme using an $(n-s)\times n$ Gaussian random matrix can be directly expressed by substituting $m+s$ by $n$ and $m$ by $n-s$ in Theorem~\ref{thm:straggler-threshold-kappa}. 
In Table~\ref{tab:threshold-comparison}, we compare the straggler threshold $s_{\kappa}$ for CommFR-GC against the straggler threshold for the scheme in~\cite{YeAbbe:18}, denoted as $s_{\kappa}^{YA}$. We take $n = 60$ as a representative for small $n$ regime, and $n = 1000$ for large $n$ regime. We choose $s$ to be around $5\%$, $10\%$, and $20\%$ of $n$. For $m$, we consider two regimes, $s \geq m$ and $s \leq m$. We observe that CommFR-GC yields much larger straggler threshold under the required numerical stability.  We have carried out simulations for a wide range of values of $n$, $s$, $m$, $\kappa$, and $\epsilon$, and observed that CommFR-GC consistently outperforms the Ye-Abbe scheme~\cite{YeAbbe:18}. 

\section{Experiments on Amazon EC2}
\label{sec:experiments}



In this section, we evaluate CommFR-GC on Amazon EC2, and compare its performance with the other schemes in the literature. 
Specifically, we compare CommFR-GC using a systematic MDS code against: (1) the na\"ive scheme, where the data is partitioned among all workers, and the server waits for all workers; 
(2) the ignoring Stragglers approach, where the data is partitioned among all workers, and the server waits for the first $n - s$ workers; and
(3) the gradient coding schemes in~\cite{Tandon:17,YeAbbe:18}. 

We train a logistic regression model on the Amazon Employee Access dataset from Kaggle, and we compare the average running time and generalization AUC (area under the curve) on a validation set. We used 26,210 training samples, and a model dimension of 241,915 (after one-hot encoding with interaction terms), and adopted Nesterov's Accelerated Gradient (NAG) descent with a constant learning rate, chosen using cross-validation.

We used Python with \texttt{mpi4py} package (similar to~\cite{Tandon:17,Raviv:18,YeAbbe:18}), \texttt{t2.micro} instances on Amazon EC2 as workers, and a single \texttt{c3.8xlarge} instance as the server. The results for $n=60$ workers are shown in Figures~\ref{fig:avg-time-n60} and~\ref{fig:AUC-n60}.
We observe that 
CommFR-GC achieves the smaller mean time per iteration by $16.02\%$ than other codes. 
Note that, the curve corresponding to CommFR-GC is always on the left side of the curves corresponding to the other schemes, which means that CommFR-CC achieves the target generalization error faster than the other schemes.

\bibliographystyle{IEEEtran}
\bibliography{Communication_Efficient_Gradient_Coding}


\appendices

\section{Coding Schemes of~\cite{YeAbbe:18}}
\label{app:YeAbbe}
\noindent {\it 1. Placement:} The schemes proposed in~\cite{YeAbbe:18} use cyclic placement, which assumes $k = n$, and assigns to the $i$-th worker the following $l$ datasets $\{D_i, D_{(i + 1) \mod n}, \ldots, D_{(i + l - 1) \mod n}\}$. 

\noindent {\it 2. Encoding:} The encoding matrix is computed as $G = BV$, where $B$ is an $mn \times (n-s)$ matrix $B$ and $V$ is an $(n-s)\times n$ matrix. The matrix $V$ must have the property that any of its $(n-s)\times(n-s)$ sub-matrix is non-singular. Further, the matrix $B$ must be such that a specific set of its rows are orthogonal to $V$ (see~\cite{YeAbbe:18} for details). 
The $i$-th worker arranges its partial gradients in a $\lceil d/m \rceil \times mn$ matrix $g_{mat}$, and computes $g_{mat}G_i$, where $G_i$ is the $i$-th column of $G$. The orthogonality property ensures that the $i$-column of $G$ has a support corresponding to the partial gradients assigned to the $i$-th worker. 

\noindent{\it 3. Decoding:} To recover the gradient sum, the server effectively needs to invert an $(n-s)\times(n-s)$ matrix $V_T$, where $T\subseteq[n]$ is the set of non-straggling workers and and $V_T$ is a $(n-s)\times |T|$ submatrix of a $(n-s)\times n$ matrix $V$ corresponding to columns indexed by $T$.

The authors present two coding schemes---when $V$ is a Vandermonde matrix, and when $V$ is a random Gaussian matrix. 


\section{Proof of Theorem~\ref{thm:straggler-threshold-kappa}}
\label{app:proof-numerical-stability}
The proof leverages the following result on the condition number of a random Gaussian matrix from~\cite[Theorem 4.5]{ChenD:05}.
\begin{lemma}(\cite{ChenD:05})
\label{thm:bound-prob-cond-number}
For any $u \geq 2$, $v\geq 2$, and $x \geq |v - u | +1$, the condition number of $u\times v$ matrix $M$ with i.i.d. standard normal elements satisfies
\begin{equation}
    \label{eq:bound-prob-cond-number}
    Pr\left(\frac{\cond{M}}{v/(|v-u|+1)} > x \right) \leq 
    \sqrt{2\pi}\left(\frac{C}{x}\right)^{|v - u| + 1},
\end{equation}
where $\cond{M}$ denotes the condition number of $M$, and $C\leq 6.414$ is a universal positive constant independent of $u$, $v$, and $\kappa$.
\end{lemma}
For a subset $T\subseteq[s+m]$, let $G_T$ denote the submatrix of $G$ consisting of columns of $G$ indexed by $T$. Suppose the server waits for the first $t$ workers from each group, where $m\leq t\leq s+m$. First, note that if  
\begin{equation}
\label{eq:subset-cond-no}
\max_{T\subseteq[s+m],|T|=t}\cond{G_T}\leq \kappa,
\end{equation}
then $s_{\kappa} \geq s+m-t$
Next, we show that~\eqref{eq:subset-cond-no} is satisfied with probability at least $1 - \epsilon$, if $\funct{s,m,\kappa}{\epsilon}\leq \epsilon$. Towards this, consider the probability that the maximum condition number over submatrices is greater than $\kappa$ as follows:
\begin{IEEEeqnarray}{l}
\Pr\left(\max_{T\subseteq[s+m],|T|=t}\cond{G_T} > \kappa\right)\\ 
\qquad \stackrel{(a)}{\leq} \binom{s+m}{t} \Pr\left(\cond{G_T}> \kappa\right)\\
\qquad \stackrel{(b)}{<}  \binom{s+m}{t}\frac{1}{\sqrt{2\pi}}\left(\frac{6.414t}{\kappa(t-m+1)}\right)^{t-m+1}\\
\qquad \stackrel{(c)}{=}  \funct{m+s,m,\kappa}{t},\\
\qquad \stackrel{(d)}{\leq} \epsilon, \quad \forall\:t\geq t^*
\end{IEEEeqnarray}
where (a) follows from the union bound, (b) follows from~\eqref{eq:bound-prob-cond-number},\footnote{Since the condition on $\kappa$ in~\eqref{eq:kappa-lower-bound} ensures that $\kappa > t-m+1$ for any $C\geq 3$, the conditions of Theorem~\ref{thm:bound-prob-cond-number} hold.} (c) follows from~\eqref{eq:function-f}, and (d) follows from the definition of $t^*$. Therefore, if $\funct{m+s,m,\kappa}{t}\leq \epsilon$, ~\eqref{eq:subset-cond-no} is satisfied with probability at least $1-\epsilon$.


\end{document}